\documentclass[a4paper,twocolumn]{article}
\usepackage[dvips]{graphicx}
\begin{document}

\title{Nonclassical correlations between photon number and quadrature
components of the light field}

\author{Holger F. Hofmann\\Department of Physics, Faculty of Science, 
University of Tokyo\\7-3-1 Hongo, Bunkyo-ku, Tokyo113-0033, Japan\\
Tel:03-5481-4228, Fax:03-5481-4240\\ 
e-mail: hofmann@femto.phys.s.u-tokyo.ac.jp}

\date{}

\maketitle

\begin{abstract}
Finite resolution quantum nondemolition (QND) measurements allow a
determination of light field properties while preserving some of the 
original quantum coherence of the input state. It is thus possible to
measure correlations between the photon number and a quadrature 
component of the same light field mode. Nonclassical features emerge
as photon number quantization is resolved. In particular, a strong
anti-correlation of quantization and coherence is observable in QND 
measurements of photon number, and a correlation between measurement
induced quantum jumps and quadrature component measurement results is
obtained in QND measurements of quadrature fluctuations in the photon
vacuum. Such nonclassical correlations represent fundamental quantum
properties of the light field and may provide new insights into the 
nature of quantization itself.
\end{abstract}

\section{Photon number and quadrature components}
When Max Planck introduced the concept of quantization one hundred
years ago \cite{Pla00}, he was painfully aware that this theory did not fit in
with Maxwell's highly successful theory of electromagnetic radiation. 
Specifically, the assumption that the energy of the continuous
light field can only have a discrete set of values appears to be 
in contradiction with the necessary continuity of interference phenomena.
In the more complete formalism of quantum mechanical operators and states,
this strange relationship of a discrete light field intensity given by
a photon number $\hat{n}$ and the continuous quadrature components 
$\hat{x}$ and $\hat{y}$ is expressed by the operator equation
\begin{equation}
\label{eq:opsum}
\hat{n}+\frac{1}{2} = \hat{x}^2 + \hat{y}^2.
\end{equation}
If $\hat{n}$, $\hat{x}$, and $\hat{y}$ were given by real numbers,
continuous shifts in $\hat{x}$ and $\hat{y}$ such as the ones caused
by interference with another field mode should cause continuous
changes in $\hat{n}$. However, $\hat{n}$, $\hat{x}$, and $\hat{y}$ 
are operators with non orthogonal eigenstates. Precise knowledge of
the eigenvalue of $\hat{n}$ therefore restricts the possible knowledge 
about $\hat{x}$ and $\hat{y}$. This uncertainty between the photon number
$\hat{n}$ and the quadrature components $\hat{x}$ and $\hat{y}$ is
a necessary requirement for the discreteness of the eigenvalues of
the photon number $\hat{n}$. 

However, uncertainty cannot explain the quantization of photon
number. Even arbitrarily small changes in $\hat{x}$ and $\hat{y}$ cause 
either no change in photon number, or seemingly discontinuous 
``quantum jumps'' changing the photon number by at least one full photon. 
The randomness of these quantum jumps may seem to defy further 
analysis. Yet, equation (\ref{eq:opsum}) indicates some correlation 
between the photon number $\hat{n}$ and the quadrature
components $\hat{x}$ and $\hat{y}$. In order to investigate this 
correlation, it is useful to consider quantum nondemolition (QND) 
measurements with a finite measurement resolution 
\cite{Cav80,Lev86,Fri92,Bru90,Hol91,Yur85,Por89,Per94}. 
Since the measured light field 
is not absorbed in a QND measurement, further measurements performed 
on a different property of the same field are possible \cite{Imo85,Kit87}. 
In this manner, photon number measurements may be combined with 
quadrature component measurements. Even though the measurement 
interaction introduces some noise into the field, the finite
measurement resolution permits a limitation of noise to the minimum
required by the uncertainty relations. The measurement result of the 
QND measurement can then be correlated with the outcome of the final 
measurement performed on the same light field mode 
\cite{Hof00a,Hof00b}. Since the noise introduced in the 
QND measurement should not depend on the measurement
result, this correlation reveals fundamental quantum mechanical 
properties of the original light field state. 

\section{General properties of QND measurements}
Optical quantum nondemolition measurements probe the light field by 
nonlinear interactions between a meter field $M$ and the signal field
$S$. Usually, the meter field is initially in a coherent state and
the quadrature component $\hat{x}_M$ of the meter field serves as 
pointer variable. The interaction between the QND variable $\hat{A}_S$
of the system and the meter components $\hat{x}_M$ and $\hat{y}_M$ is
then given by
\begin{eqnarray}
\hat{x}_M(\mbox{out}) &=& \hat{x}_M(\mbox{in})
             +\frac{\hat{A}_S(\mbox{in})}{2\delta\!A}
\nonumber \\
\hat{y}_M(\mbox{out}) &=& \hat{y}_M(\mbox{in})
\nonumber \\[0.3cm]
\hat{A}_S(\mbox{in})  &=& \hat{A}_S(\mbox{in}).
\end{eqnarray}
Note that the measurement resolution $\delta\! A$ is a function of the
initial meter fluctuation of $\delta\!x_M=1/2$ and the coupling strength 
provided by the nonlinearity. Signal variables $\hat{B}_S$ which do not
commute with the QND variable $\hat{A}_S$ are subject to random changes
induced by the quantum fluctuations of $\hat{y}_M$ according to  
\begin{eqnarray}
\hat{B}_S(\mbox{out}) &=& \hat{U}^{-1}\hat{B}_S(\mbox{in})\hat{U}
\nonumber \\[0.3cm]
\mbox{with}\hspace{0.3cm}\hat{U}&=&
\exp\left(-i\frac{\hat{A}_S(\mbox{in})
               \; \hat{y}_M(\mbox{in})}{\delta\!A}\right).
\end{eqnarray}
After the interaction, the meter field and the signal field are entangled.
A measurement of $\hat{A}_S$ can still be avoided by reading out the noise 
variable $\hat{y}_M$. The change of the system is then found to be a 
unitary transformation and no information about system properties is
obtained. If the meter variable $\hat{x}_M$ is read out, however, information
about the system variable $\hat{A}_S$ is obtained while the uncertainty of
the noise term $\hat{y}_M$ causes an uncontrollable change in all other 
system properties $\hat{B}_S$. 

By identifying the measurement readout of $\hat{x}_M$ directly with the 
most likely value $A_m$ of $\hat{A}_S$, the measurement can be represented
by a generalized measurement operator $\hat{P}_{\delta\! A}(A_m)$ given by
\begin{equation}
\label{eq:post}
\hat{P}_{\delta\!A}(A_m) = (2\pi \delta\! A^2)^{-1/4}\; 
              \exp \left(-\frac{(\hat{A}_S-A_m)^2}{4 \delta\! A^2}\right).
\end{equation}
Note that the measurement values $A_m$ are continuous even if $\hat{A}_S$ 
has only discrete eigenvalues. In this sense, the generalized measurement
operator $\hat{P}_{\delta\! A}(A_m)$ overcomes the limitation to 
eigenvalues inherent in the conventional 
measurement postulate \cite{Hof00c}.
The whole effect of a measurement of $A_m$ with a resolution $\delta\! A$
can now be described by the operator $\hat{P}_{\delta\! A}(A_m)$. 
For an initial state $\mid \psi_S(\mbox{in}) \rangle$ of the signal field, 
the probability distribution $P(A_m)$ over measurement
results $A_m$ and the state $\mid \psi_S (A_m) \rangle$
conditioned by a measurement result of $A_m$ are given by
\begin{eqnarray}
P(A_m) &=& \langle \psi_S(\mbox{in}) \mid \hat{P}^2_{\delta\!A}(A_m)
\mid \psi_S(\mbox{in}) \rangle
\nonumber \\
\mid \psi_S (A_m) \rangle &=& \frac{1}{\sqrt{P(A_m)}}
          \hat{P}_{\delta\!A}(A_m)\mid \psi_S(\mbox{in}) \rangle.
\end{eqnarray}
It is then possible to derive correlations between the measurement result
$A_m$ and further measurements by referring to the statistical properties
of the conditioned output state $\mid \psi_S (A_m) \rangle$.

\section{Anti-correlation of quantization and coherence}
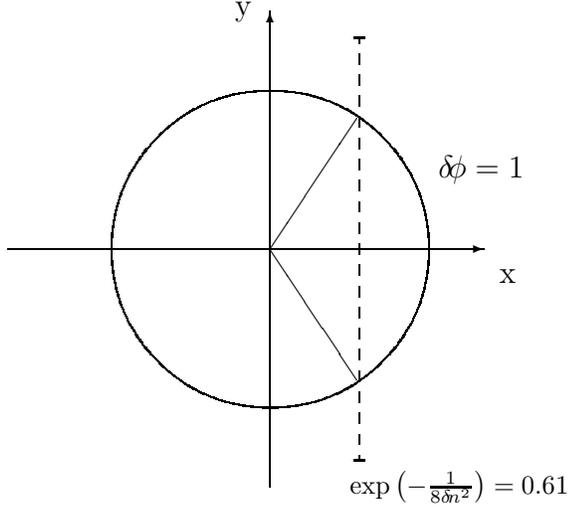
\begin{figure}[hbt]
\begin{picture}(240,200)
\put(110,10){\vector(0,1){180}}
\put(11,100){\vector(1,0){180}}
\put(90,180){\makebox(20,20){\large y}}
\put(190,80){\makebox(20,20){\large x}}
\bezier{100}(110,40)(135,40)(152.5,57.5)
\bezier{100}(152.5,57.5)(170,75)(170,100)
\bezier{100}(170,100)(170,125)(152.5,142.5)
\bezier{100}(152.5,142.5)(135,160)(110,160)
\bezier{100}(110,40)(85,40)(67.5,57.5)
\bezier{100}(67.5,57.5)(50,75)(50,100)
\bezier{100}(50,100)(50,125)(67.5,142.5)
\bezier{100}(67.5,142.5)(85,160)(110,160)
\put(110,100){\line(2,3){33}}
\put(110,100){\line(2,-3){33}}
\put(180,120){\makebox(20,20){\large $\delta\!\phi=1$}}
\put(130,0){\makebox(100,20){
$\exp\left(-\frac{1}{8\delta\!n^2}\right)=0.61$}}
\put(144,20){\dashbox{4}(0,160)}
\end{picture}
\caption{\label{nphi} \small Illustration of the relationship between phase
randomization and decoherence in the expectation value 
$\langle \hat{a} \rangle$ for a photon number measurement resolution 
of $\delta\! n = 0.5$.}
\end{figure}

QND measurements of photon number have been realized 
experimentally using fiber optics \cite{Lev86,Fri92}. In these setups,
a third order nonlinearity shifts the phase of the coherent meter field 
by an amount proportional to the intensity of the signal field,
while the intensity fluctuations of the meter field cause a randomization
of the signal phase according to the uncertainty relation
\begin{equation}
\delta\!\phi = \frac{1}{2 \delta\! n}.
\end{equation}
The expectation value of the signal field amplitude 
$\hat{a}=\hat{x}+i\hat{y}$ is consequently reduced to
\begin{equation}
\langle \hat{a} \rangle_{\mbox{out}} = \exp\left(-\frac{1}{8\delta\!n^2}\right)
\langle \hat{a} \rangle_{\mbox{in}},
\end{equation}
as illustrated in figure \ref{nphi}. 

\begin{figure}[hbt]
\begin{picture}(240,270)
\put(20,150){\makebox(200,100){
\includegraphics[width=6cm]{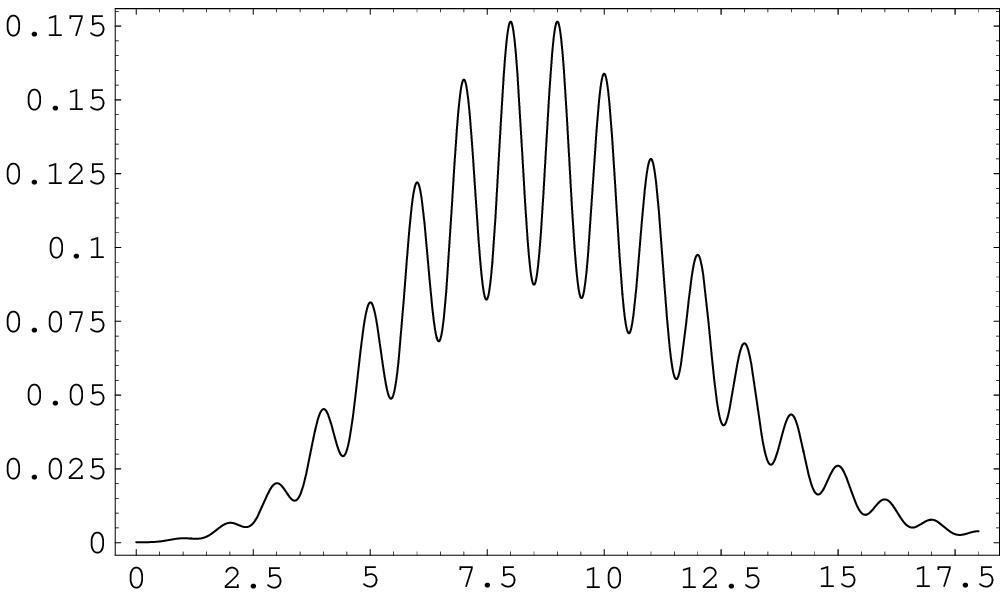}}}
\put(60,225){\makebox(20,20){(a)}}
\put(120,130){\makebox(20,20){$n_m$}}
\put(0,200){\makebox(40,20){$P(n_m)$}}
\put(24,30){\makebox(200,100){
\includegraphics[width=6cm]{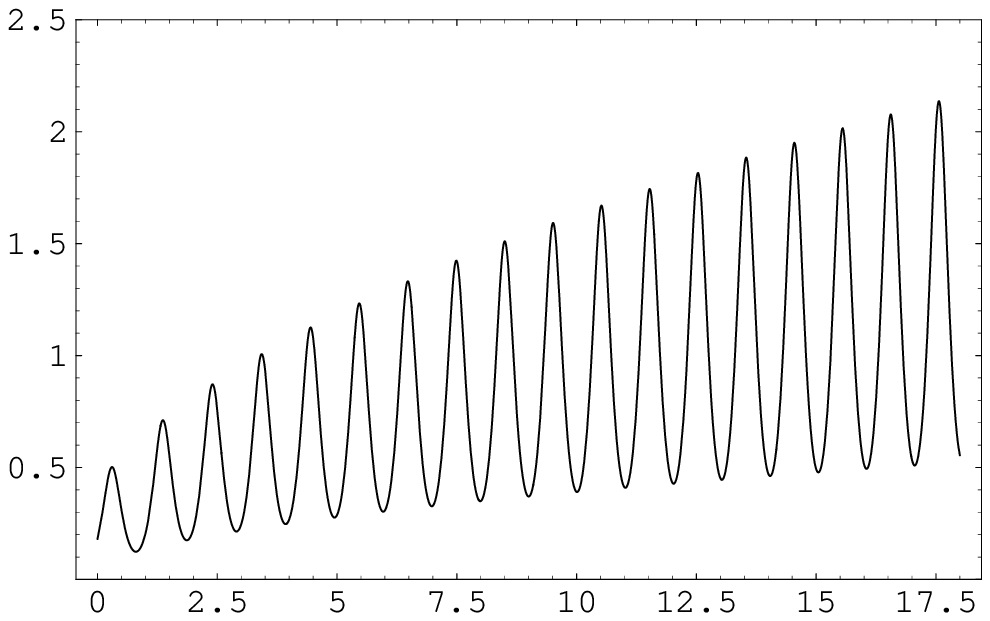}}}
\put(60,105){\makebox(20,20){(b)}}
\put(120,10){\makebox(20,20){$n_m$}}
\put(0,70){\makebox(50,20){$\langle\hat{a}\rangle_f(n_m)$}}
\end{picture}
\caption{\label{phcorr} \small Measurement statistics for a coherent 
state amplitude of $\alpha=3$ and a resolution of $\delta\! n=0.3$.
(a) shows the photon number measurement probabilities $P(n_m)$
and (b) the coherence $\langle\hat{a}\rangle_f (n_m)$ after
a measurement result of $n_m$.}
\end{figure}

These dephasing characteristics have been discussed previously and
correspond well with the measurement dynamics observed experimentally
\cite{Imo85,Kit87}. However, the averaged results hide a peculiar
correlation between the dephasing statistics and the measurement results
which can be obtained by applying the generalized measurement operator
\cite{Hof00a}.
If the initial signal field is in a coherent state $\mid \alpha \rangle$
given by 
\begin{equation}
\mid \alpha \rangle = \exp(-\frac{|\alpha|^2}{2}) \sum_n 
        \frac{\alpha^n}{\sqrt{n!}} \mid n \rangle,
\end{equation}
the probability distribution over measurement results $n_m$ is given
by
\begin{eqnarray}
\label{eq:xPresult}
P(n_m) &=& \langle \alpha \mid \hat{P}_{\delta\! n}^2(n_m) \mid \alpha \rangle
\nonumber \\[0.3cm] &=& \frac{\exp(-|\alpha|^2)}{\sqrt{2\pi\delta\! n^2}}
\nonumber \\ && \times
      \sum_n \frac{|\alpha|^{2n}}{n!} 
             \exp\left(-\frac{(n-n_m)^2}{2\delta\! n^2}\right),
\end{eqnarray}
and the expectation value of the coherent amplitude $\hat{a}$ after
the measurement is given by
\begin{eqnarray}
\label{eq:xCresult}
\langle \hat{a} \rangle_f(n_m) &=& 
\frac{\langle \alpha \mid \hat{P}_{\delta\! n}(n_m)
             \hat{a} \hat{P}_{\delta\! n}(n_m) \mid \alpha \rangle}
     {\langle \alpha \mid \hat{P}_{\delta\! n}^2(n_m) \mid \alpha \rangle}
\nonumber \\
&=& \alpha \; \exp\left(-\frac{1}{8\delta\! n^2}\right)
\nonumber \\ && \times
\frac{\sum_n \frac{|\alpha|^{2n}}{n!} 
             \exp\left(-\frac{(n+\frac{1}{2}-n_m)^2}{2\delta\! n^2}\right)}
     {\sum_n \frac{|\alpha|^{2n}}{n!} 
             \exp\left(-\frac{(n-n_m)^2}{2\delta\! n^2}\right)}.
\end{eqnarray}
The results for $\alpha=3$ and a resolution of $\delta\! n=0.3$ are
shown in figure \ref{phcorr}.
After the measurement, the expectation values of the coherent amplitude
$\hat{a}$ are maximal if the measurement result $n_m$ was a half integer 
value and minimal if it was an integer value. Therefore, the accidental
measurement of a properly quantized integer photon number causes additional
phase noise, while the accidental observation of a half-integer photon number
preserves the original phase coherence of the field. 

\begin{figure}[hbt]
\begin{picture}(240,150)
\put(30,20){\makebox(200,100){
\includegraphics[width=6cm]{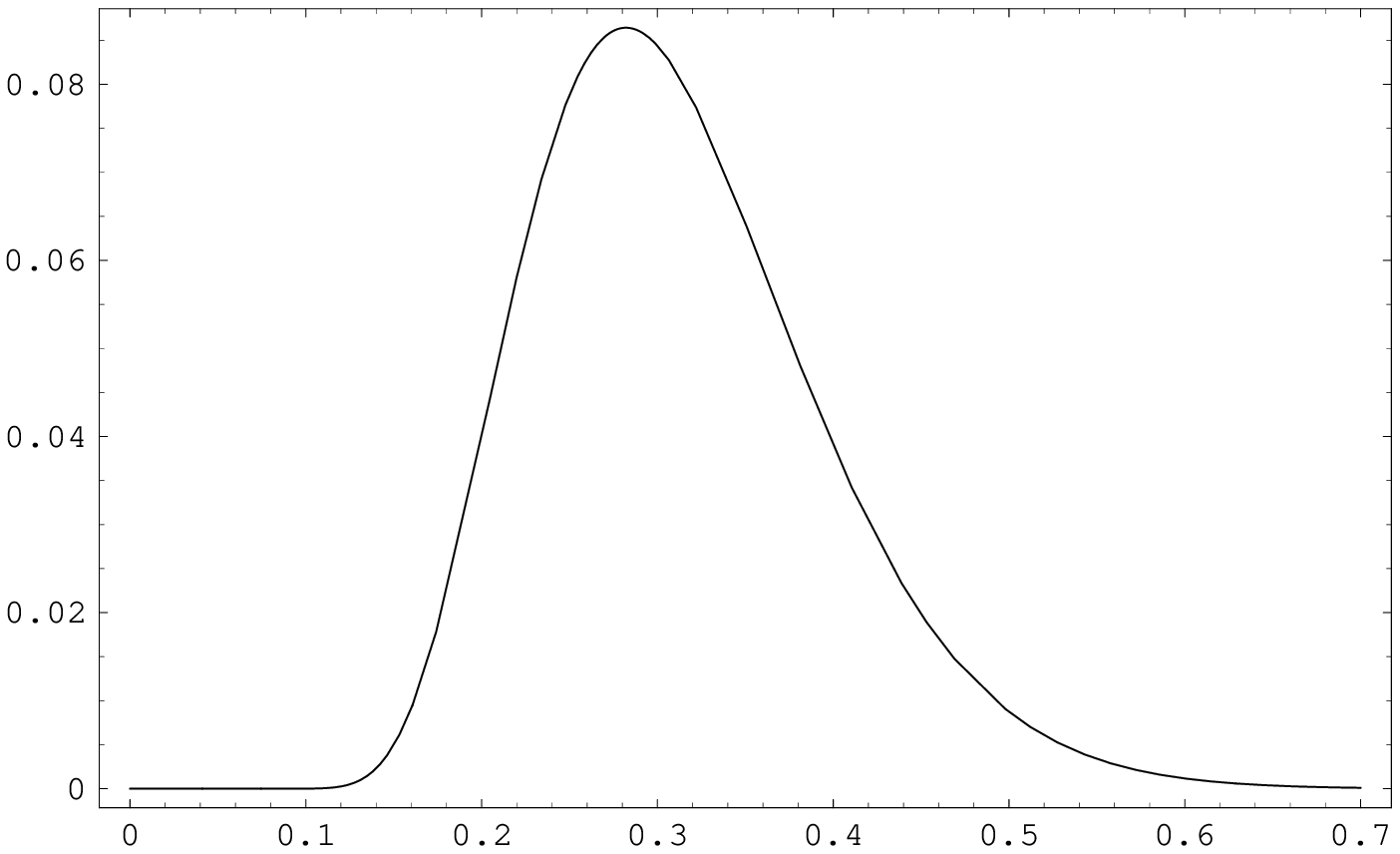}}}
\put(120,0){\makebox(20,20){\large $\delta\! n$}}
\put(0,70){\makebox(50,20){$-\frac{C(Q;\langle\hat{a}\rangle_f)}{\alpha}$}}
\end{picture}
\caption{\label{qacorr} \small Anti-correlation between the quantization 
Q and the
coherence $\langle\hat{a}\rangle_f$ after the measurement as a function of
measurement resolution $\delta\!n$.}
\end{figure}

In order to quantify this property, it is useful to define the 
quantization $Q$ of the measurement result $n_m$ as
\begin{equation}
Q(n_m)=\cos\left(2\pi n_m\right).
\end{equation}
It is then possible to determine the correlation 
$C(Q;\langle \hat{a}\rangle_f)$ between quantization $Q$
and coherence $\langle\hat{a}\rangle_f$ by
\begin{eqnarray}
\label{eq:corr}
\lefteqn{C(Q;\langle \hat{a}\rangle_f) =} \nonumber \\[0.3cm] &&
 \overline{Q \; \langle \hat{a}\rangle_f} 
- \overline{Q} \; \overline{\langle \hat{a}\rangle_f} 
\nonumber \\
&=& - 2 \exp\left(-2\pi^2\delta\! n^2\right) 
        \exp\left(-\frac{1}{8 \delta\! n^2}\right) \alpha,
\end{eqnarray}
where the over-lined quantities are averages over all possible measurement
values $n_m$. The resulting correlation is always negative, since the
coherence is maximal at half-integer photon numbers which have a 
quantization $Q$ of minus one. Figure \ref{qacorr} shows the dependence 
of this anti-correlation between quantization and coherence as a
function of measurement resolution.
At resolutions $\delta\!n\gg 0.3$, there is no correlation because
quantization is not resolved. At resolutions $\delta\!n\ll 0.3$,
there is no correlation because the phase is completely randomized and the
average coherence $\langle\hat{a}\rangle_f$ is reduced to zero. In the
intermediate regime, however, quantization and coherence are clearly 
anti-correlated properties of the light field. 

While the correlation $C(Q;\langle\hat{a}\rangle_f)$ between quantization
and coherence is definitely an observable property of the light field, 
an experimental verification in the optical region is difficult 
because of the relative weakness of the available nonlinearities. In the QND 
measurements using fiber optics, the resolutions achieved are still far 
below the quantum limit (e.g. 
$\delta\! n \approx 10^{-4}$ in \cite{Fri92}). Alternatively, it is possible
to investigate nonclassical correlations by first realizing a QND measurement
of coherent field properties, followed by a precise photon number measurement.

\section{Correlation of field fluctuations and quantum jumps}
\begin{figure}[hbt]
\begin{picture}(240,150)
\put(30,20){\makebox(200,100){
\includegraphics[width=6cm]{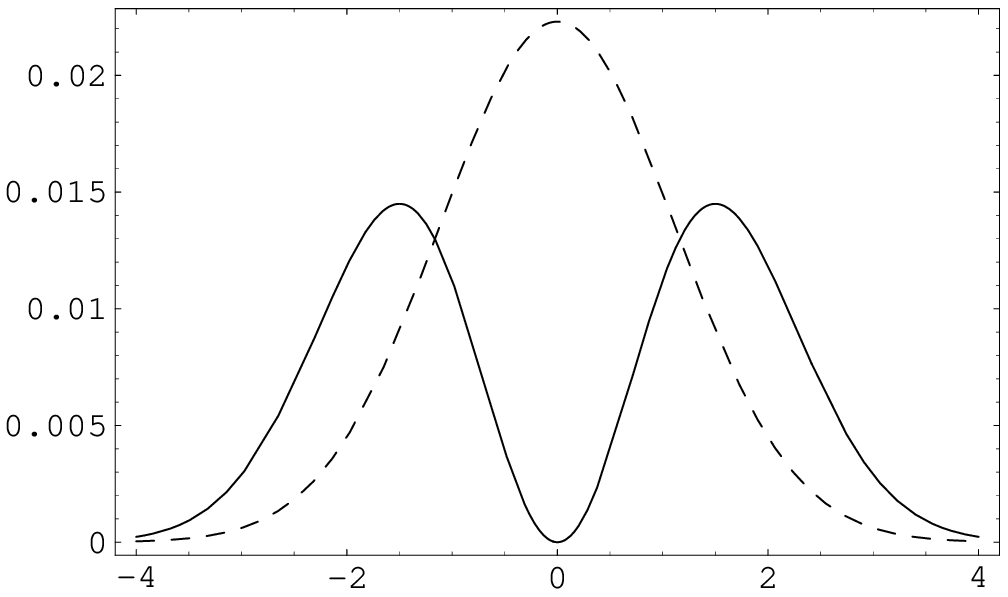}}}
\put(140,0){\makebox(20,20){\large $x_m$}}
\put(0,70){\makebox(50,20){$P_1(x_m),$}}
\put(0,50){\makebox(50,20){$\frac{P(x_m)}{16}$}}
\end{picture}
\caption{\label{qjump} \small Conditional probability $P_1(x_m)$ at a 
resolution of $\delta\!x =1$. The dashed line shows the total Gaussian 
probability $P(x_m)$, reduced by a factor of 1/16 for comparison.}
\end{figure}

QND measurements of quadrature components have been realized using the 
phase sensitive nonlinear interaction in optical parametric amplifiers (OPAs) 
\cite{Yur85,Por89,Per94}. By exploiting the interference properties
of two phase sensitive amplification steps, it is possible to shift one
quadrature component of the meter field by an amount proportional to
the signal field component $\hat{x}$. 
\begin{figure}
\begin{picture}(240,200)
\put(120,0){\vector(0,1){200}}
\put(30,100){\vector(1,0){180}}
\put(80,180){\makebox(20,20){y}}
\put(200,80){\makebox(20,20){x}}
\put(100,80){\makebox(20,20){$0$}}
\put(115,130){\line(1,0){5}}
\put(150,95){\line(0,1){5}}
\put(140,75){\makebox(20,20){$\frac{1}{4}$}}
\put(95,120){\makebox(20,20){$\frac{1}{4}$}}
\bezier{100}(120,40)(145,40)(162.5,57.5)
\bezier{100}(162.5,57.5)(180,75)(180,100)
\bezier{100}(180,100)(180,125)(162.5,142.5)
\bezier{100}(162.5,142.5)(145,160)(120,160)
\bezier{100}(120,40)(95,40)(77.5,57.5)
\bezier{100}(77.5,57.5)(60,75)(60,100)
\bezier{100}(60,100)(60,125)(77.5,142.5)
\bezier{100}(77.5,142.5)(95,160)(120,160)
\bezier{100}(120,15)(145,15)(162.5,40)
\bezier{100}(162.5,40)(180,65)(180,100)
\bezier{100}(180,100)(180,135)(162.5,160)
\bezier{100}(162.5,160)(145,185)(120,185)
\bezier{100}(120,15)(95,15)(77.5,40)
\bezier{100}(77.5,40)(60,65)(60,100)
\bezier{100}(60,100)(60,135)(77.5,160)
\bezier{100}(77.5,160)(95,185)(120,185)
\end{picture}
\caption{\label{squeez} \small Illustration of the change in the quantum noise 
distribution due to a measurement of $\hat{x}$ with a resolution of 
$\delta\!x=0.5$. The circle represents vacuum noise (1/2 in each component),
the ellipse represents the increased noise after the measurement interaction.
}
\end{figure}
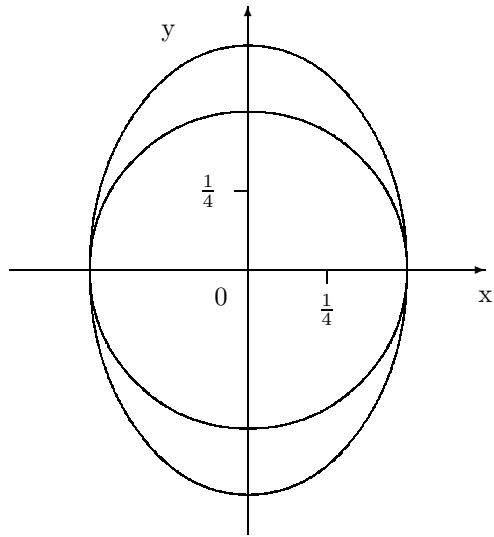
Since the uncertainty relation between
the quadrature components is given by
\begin{equation}
\delta\! x\delta\!y \geq \frac{1}{4},
\end{equation}
this measurement increases the noise in the $\hat{y}$ component as
illustrated in figure \ref{squeez}. While there appears to be no
correlation between the measurement result $x_m$ and the change in
$\hat{y}$, it is possible to establish a connection between the 
measurement result $x_m$ and changes in the photon number $\hat{n}$.
The natural input state for this investigation is the vacuum field
$\mid 0 \rangle$ with its well defined photon number of zero.
The total probability distribution $P(x_m)$ over measurement results
$x_m$ is then given by
\begin{eqnarray}
\lefteqn{P(x_m)=|\langle 0 \mid \hat{P}^2_{\delta\!x}(x_m) \mid 0 \rangle|}
\nonumber \\
      &=& \sqrt{\frac{2}{\pi(1+4\delta\!x^2)}} 
          \exp\left(-\frac{2 x_m^2}{1+4\delta\!x^2}\right).
\end{eqnarray}
For high resolutions ($\delta\! x\to 0$), this Gaussian distribution 
reproduces the quantum noise level of $\langle \hat{x}^2 \rangle = 1/4$. 
At low resolutions ($\delta\! x > 1/2$), the measurement uncertainty
dominates. The measurement induced changes in the quantum state of the 
signal field are described by $\hat{P}_{\delta\!x}(x_m)$. At low 
resolution, this change is small and the vacuum component is still dominant 
in the output field. However, a photon counting measurement in the signal
output analyzes this slight change in terms of quantum jumps from zero
to one photon. The joint probability $P_1(x_m)$ of measuring $x_m$ and 
observing a photon in the output signal is given by
\begin{eqnarray}
\lefteqn{
P_1(x_m) = |\langle 1 \mid \hat{P}_{\delta\!x}(x_m) \mid 0 \rangle|^2}
\nonumber \\[0.1cm]
&=& \frac{1}{\sqrt{2\pi\delta\!x^2}} 
\frac{32\; \delta\!x^2}{(1+8 \delta\!x^2)^3} x_m^2 
\exp\left(-\frac{4 x_m^2}{1+8 \delta\!x^2}\right).
\nonumber \\
\end{eqnarray}
For low resolutions, 
this symmetric, double peaked probability distribution has its maxima near 
$x_m=\pm \sqrt{2}\; \delta\!x$. A comparison of $P_1(x_m)$ 
and $P(x_m)$ at a resolution of $\delta\! x=1$ is shown in figure \ref{qjump}.
Note that the total probability shown in figure \ref{qjump} is reduced 
by $1/16$, since the total probability of observing a photon in the output 
at $\delta\!x=1$ is equal 
to $1/16$. The double peaked distribution has its peaks at the flanks of the
total distribution, indicating that quantum jump events are always
associated with high field fluctuations. The photon number after the 
measurement is therefore correlated with the measurement result. This 
correlation can be written as
\begin{eqnarray}
\lefteqn{C(x_m^2;\langle \hat{n}\rangle_f) =} \nonumber \\ && 
\int \langle 0 \mid \hat{P}_{\delta\!x}(x_m)\hat{n}\hat{P}_{\delta\!x}(x_m) 
\mid 0 \rangle \; x_m^2 \, d\!x_m.
\end{eqnarray}
The integral over the measurement results $x_m$ and the expectation 
values of $\hat{n}$ after the measurement can be solved by making use of the
operator properties given by equation (\ref{eq:post}). It can then 
be writtten as an
operator correlation which does not depend on the measurement resolution,
\begin{eqnarray}
\label{eq:nonclass}
\lefteqn{C(x_m^2;\langle \hat{n}\rangle_f) =} \nonumber \\ &&
\frac{1}{4}\langle \hat{x}^2\hat{n}+2\hat{x}\hat{n}\hat{x}+\hat{n}\hat{x}^2
\rangle - \langle \hat{x}^2\rangle\langle\hat{n}\rangle \;=\;\frac{1}{8}. 
\end{eqnarray}
The characteristics of the quantum jump statistics illustrated by figure
\ref{qjump} are therefore directly related to fundamental properties of the
operator formalism. In particular, equation (\ref{eq:nonclass}) shows that
an operator correlation between photon number and field fluctuations is 
possible even if the field is in a photon number eigenstate. This correlation
is a direct consequence of the non-commutativity of operators, since the
sandwiching of the photon number operator $\hat{n}$ between the field
operators $\hat{x}$ makes the eigenvalue of $n=0$ in the photon number state
irrelevant. Indeed, the photon number of the vacuum is only zero with respect
to actual photon number measurements. It cannot be considered a measurement
independent physical property of the system. Much of
the confusion surrounding the interpretation of quantum mechanics arises from 
an erroneous identification of eigenvalues with such measurement independent
``elements of reality''.

\section{Implications for the interpretation of quantization}
Both the anti-correlation of quantization and coherence and the correlation
of field fluctuations and quantum jumps indicate that the discreteness of
photon number is not an intrinsic property of the light field itself but
a property of the specific measurement interaction. If the photon number 
is not resolved, it should not be considered an integer number. Eigenvalues
of operator variables do not represent the ``real'' physical values of that 
property. The EPR paradox\cite{EPR} and Bell's inequalities\cite{Bel64}
clearly illustrate the fallacy of attempting an identification of
eigenvalues with ``elements of reality''. In particular, there is every
reason to reject the assumption that ``{\it If, without in any way 
disturbing a system, we can predict with certainty the value of a 
physical quantity, then there exists an element of physical reality
corresponding to this physical quantity.}''\cite{EPR}. If the predicted
experiment is never performed, it is pointless to demand an ``element of
reality'' for something that might have been. The anti-correlation of
quantization and coherence shows that half integer photon numbers are
a relevant part of the correlated photon-field statistics.
The quantum jump correlation shows that the photon number of the vacuum 
is effectively nonzero when field measurements are performed first. 
The nature of the quantum mechanical formalism itself thus demands a 
dependence of reality on the actual measurement situation. 
Planck's problem of reconciling the discreteness of photon number with 
the continuity of interference in the light field can only be overcome 
by admitting the context dependence of quantum mechanical reality 
expressed by the operator formalism. An operator does not represent a 
numerical value. Rather, it represents a potential interaction with  
its environment. Instead of abstractly analyzing states and eigenvalues, 
it is therefore necessary to explore quantum mechanical properties from 
the perspective of a realistic measurement context. 

\section*{Acknowledgements}
I would like to acknowledge support from the Japanese 
Society for the Promotion of Science, JSPS.



\end{document}